\documentclass[%
pre,
tightenlines,
showpacs,showkeys,
a4paper,12pt
]{revtex4}
\pdfoutput=1
\usepackage{dcolumn}
\usepackage{amssymb,amsmath,stmaryrd,array}
\usepackage{graphicx}
\usepackage{subfigure}

\renewcommand{\thesubfigure}{(\alph{subfigure})}

\makeatletter
\renewcommand{\@thesubfigure}{\thesubfigure\space}

\newcommand{\bs}[1]{\boldsymbol{#1}}
\newcommand{\vc}[1]{\mathbf{#1}}
\newcommand{\mvc}[1]{\boldsymbol{\mathcal #1}}
\newcommand{\uvc}[1]{\mathbf{\hat #1}}

\newcommand{\sca}[2]{\bigl({#1}\cdot{#2}\bigr)}

\newcommand{\cnj}[1]{{#1}^{\ast}}
\newcommand{\hcnj}[1]{{#1}^{\dagger}}

\newcommand{\pdrs}[1]{\partial_{#1}}

\newcommand{\mum}{$\mu$m}

\newcommand{\sign}{\mathop{\rm sign}\nolimits}
\newcommand{\diag}{\mathop{\rm diag}\nolimits}
\renewcommand{\Re}{\mathop{\rm Re}\nolimits}
\renewcommand{\Im}{\mathop{\rm Im}\nolimits}

\newcommand{\dd}{\mathrm{d}}
\newcommand{\ee}{\mathrm{e}}

\newcommand{\inc}{\mathrm{inc}}
\newcommand{\refl}{\mathrm{refl}}
\newcommand{\trans}{\mathrm{tr}}
\newcommand{\vac}{\mathrm{vac}}
\newcommand{\med}{\mathrm{m}}
\newcommand{\ellpt}{\mathrm{ell}}

\makeatother

\begin{document}
\DeclareGraphicsExtensions{.jpg,.png}
\title{Polarization resolved angular patterns in nematic\\
liquid crystal cells}

% \title{Polarization structure of light transmitted through nematic
% liquid crystal cells: angular patterns and singularities} 

\author{Alexei~D.~Kiselev}
\email[Email address: ]{kiselev@iop.kiev.ua}

\affiliation{%
 Institute of Physics of National Academy of Sciences of Ukraine,
 prospekt Nauki 46,
 03028 Ky\"{\i}v, Ukraine} 

\date{\today}

\begin{abstract}
We study the angular structure of polarization
of light  transmitted through a nematic liquid crystal (NLC) cell
by theoretically analyzing 
the polarization state as a function of the incidence angles.
For a uniformly aligned NLC cell, 
the $4\times 4$ matrix formalism and the orthogonality relations 
are used to derive the exact expressions for 
the transmission and reflection matrices.
The polarization resolved angular patterns in the two-dimensional
projection plane are characterized in terms of the polarization
singularities such as C points (points of circular polarization) and L lines
(lines of linear polarization).
In the case of linearly polarized plane waves incident on the
homeotropically aligned cell, we present the results of detailed
theoretical analysis describing the structure of the polarization
singularities.
We apply the theory to compute the polarization patterns
for various orientational structures in the NLC cell
and discuss the effects induced by 
director orientation and biaxiality.
\end{abstract}

 \maketitle

%%%%%%%%%%%%%%
\section{Introduction}
\label{sec:intro}
%%%%%%%%%%%%%%

The optical properties of nematic liquid crystals (NLCs)
have long been known to 
play a key part in a large variety of their 
applications~\cite{Gennes:bk:1993,Yeh:1998,Chigr:1999}.
Typically, NLCs are used as
optically anisotropic materials where 
the anisotropy is determined by
the orientational structure which is sensitive to external
fields and, in restricted geometries, 
can also be influenced by changing the boundary conditions.

Similar to other anisotropic materials, the polarization state of
light propagating through a liquid crystal varies due to the presence
of the anisotropy. When the NLC cell is placed between two crossed
polarizers, these anisotropy induced changes of the polarization
manifest themselves in variations of the intensity of the transmitted
light. This effect underlines the vast majority of experimental
techniques devised to characterize orientational structures in NLC
cells.

For example,
the crystal rotation technique~\cite{Berrem:pla:1976}
relies on analyzing the 
angle dependence of the transmittance of NLC cells
placed between crossed polarizers. 
It can be regarded as a special case
dealing with one-dimensional cross-section of the conoscopic pattern. 
% The magnetic null methods is capable of detecting hybridity
% and was initially suggested in~\cite{Sheff:jap:1977}.

Nowadays the conoscopy is proved to be a useful tool
for studying liquid crystal systems.
It is used to detect biaxiality of
NLCs~\cite{Yu:prl:1980,Madsen:prl:2004} and
to measure the pretilt angle in uniaxial liquid crystal
cells~\cite{Komit:crt:1984,Brett:apo:2001}.
Orientational structures and helix unwinding process in
ferroelectric smectic liquid crystals were also studied by conoscopy
in Refs.~\cite{Gorec:jjap:1990,Suwa:jjap:2003}.
The conoscopic patterns of 
hybrid NLC cells were investigated in Ref.~\cite{Nast:ujpo:2001}. 

In this paper we are aimed to study the polarization structure behind
the conoscopic images by performing a comprehensive analysis of the
polarization state of the light transmitted through a NLC cell
as a function of the incidence angles.
Technically, we deal with angular two-dimensional distribution of the
Stokes parameters describing the field of polarization ellipses which
might be called the 
\textit{polarization resolved conoscopic (angular) pattern}.

It was originally recognized by 
Nye~\cite{Nye:prsl:1983a,Nye:prsl:1987,Nye:bk:1999} that the important
elements characterizing the geometric structure of such polarization
fields are the so-called \textit{polarization singularities}.
In particular, these are the \textit{C points}
[the points where the light wave is circularly polarized]
and the \textit{L lines} [the curves along which the polarization is
linear].
Over the past two decades these singularities and related issues
have been the subject of numerous theoretical, experimental and numerical 
studies~\cite{Hajnal:prsl:1987a,Hajnal:prsl:1987b,Hajnal:prsl:1990,
Freund:pra:1994,Freund:optl:2002,Dennis:prsl:2000,Dennis:prsl:2001a,
Melnikov:jopb:2001,Freund:optcom:2002,Soskin:optcom:2002,Soskin:optl:2003,
Mokhun:optl:2002,Dennis:optcom:2002}.

The theory of polarization singularities was applied to
study the angular dependence of the polarization of plane wave
eigenmodes in birefringent dichroic chiral
crystals~\cite{Dennis:prsl:2003}.
The polarization state of the electric displacement field
was analyzed in relation to the direction of the wave vector.
This analysis was recently generalized and extended to 
a more complicated case of bianisotropic media~\cite{Berry:prsl:2005}.

The experimental results and theoretical analysis presented
in Ref.~\cite{Dennis:prl:2005} deal with the unfolding of
a linearly polarized Laguerre-Gauss (LG$_{01}$) beam
with an on-axis vortex
on propagation through a birefringent crystal.
It was found that 
a complicated pattern of polarization singularities is formed
as a result of  the anisotropy induced symmetry breaking.

Paraxial propagation of a beam 
carrying the angular momentum along the optic axis of a uniaxially
anisotropic medium was also studied in 
Refs.~\cite{Ciattoni:pre:2002,Ciattoni:pre:2003,Volyar:jopa:2004}.
Interestingly,
there are a number of dynamical effects
induced by a light beam
normally impinging on a homeotropically aligned NLC 
cell~\cite{Piccir:prl:2001,Piccir:pre2:2004,Piccir:pre1:2004,
Piccir:pre:2006,Marruc:prl:2006}
where liquid crystals appear as unique
anisotropic materials
sensitive to the spin and orbital angular momentum of light.

In this study we adapt a systematic theoretical approach and
explore the characteristic features of
the polarization structure
(the polarization resolved angular pattern) emerging 
due to interference of four eigenmodes excited in NLC cells
by the plane waves with varying direction of incidence.

The layout of the paper is as follows.
The problem of light transmission through a uniformly
anisotropic NLC cell is considered
in Sec.~\ref{sec:transm-light}.
Using the $4\times 4$ matrix formalism
and the orthogonality relations we deduce the exact expressions
for the transmission and reflection matrices.
In Sec.~\ref{sec:angular-struct}
the analytical results are used for analysis
of the polarization resolved conoscopic patterns
when the incident light is linearly polarized.

The case of homeotropic NLC cells
is treated in Sec.~\ref{subsec:homeot-cell}.
We show that there are concentric circles
each containing four symmetrically arranged
C points of alternating handedness.
For these C points, we obtain
the analytical expressions for the topological index
and for the discriminant whose sign determines 
the type of the point according to the line classification.
The index and the handedness of the C points
are found to alternate in sign along the radial direction.
Morphology of the C points lying on the circle of the smallest
radius representing the direction close to the normal incidence
is shown to be of the \textit{lemon} type with the index equal to $+1/2$.
In sufficiently thick cells, other C points are either of the
\textit{star} type (the index is $-1/2$) or of the \textit{monstar}
type (the index is $+1/2$).
The L lines are formed by two rotated coordinate axes
and by circles separating the circles with the C points.
Some numerical results for the tilted, planar and biaxial structures
are presented in Sec.~\ref{subsec:director-unfold}.  

% The experimental procedure is described in Sec.~\ref{sec:experiment}.
% We present the experimental results and 
% compare the predictions of the theory with the experimental data.
Discussion and concluding remarks are given in
Sec.~\ref{sec:disc-concl}.

%%%%%%%%%%%%%%%%%%%%%%%%%%%%%%%%%%%%
\section{Transmission of light through liquid crystal cells}
\label{sec:transm-light}
%%%%%%%%%%%%%%%%%%%%%%%%%%%%%%%%%%%%

We consider a nematic liquid crystal (NLC) cell of thickness $d$ 
sandwiched between two
parallel plates that are normal to the $z$ axis: $z=0$ and $z=d$.
The NLC represents a birefringent material with the dielectric tensor
given by
\begin{align}
  \label{eq:diel-tens}
  \bs{\varepsilon}=\epsilon_3\,\mvc{I}_3+\Delta\epsilon_1\,\uvc{d}\otimes\uvc{d}
+\Delta\epsilon_2\,\uvc{m}\otimes\uvc{m},
\end{align}
where $\Delta\epsilon_i=\epsilon_i-\epsilon_3$
and  
$\mvc{I}_n$ is the $n\times n$ identity matrix. 
(In what follows carets will denote unit vectors.)

Typically, anisotropy of nematics is locally uniaxial
and NLC molecules align on average along a local unit 
director~\cite{Gennes:bk:1993}.
In this case the NLC director $\uvc{d}$ determines the optic axis 
and the expression~\eqref{eq:diel-tens} taken
in the limit of vanishing biaxiality with $\Delta\epsilon_2=0$
gives the uniaxially anisotropic dielectric tensor. 
Its two principal values $\epsilon_3\equiv\epsilon_{\perp}$ and
$\epsilon_1\equiv\epsilon_{\parallel}$
define the ordinary and extraordinary refractive indices, 
$n_o=\sqrt{\mu\epsilon_{\perp}}$ and
 $n_e=\sqrt{\mu\epsilon_{\parallel}}$,
where $\mu$ is the NLC magnetic permeability.

In a more general case of biaxial 
nematics~\cite{Luckh:thsf:2001,Luckh:nat:2004}
that were recently observed
experimentally~\cite{Madsen:prl:2004,Severing:prl:2004},
there are three different dielectric constants $\epsilon_1$, $\epsilon_2$ and
$\epsilon_3$ representing the eigenvalues of the dielectric
tensor~\eqref{eq:diel-tens},
so that the eigenvectors $\uvc{d}$, $\uvc{m}$ and 
$\uvc{l}=\uvc{d}\times\uvc{m}$
give the corresponding principal axes. 
The unit vectors $\uvc{d}$, $\uvc{m}$ and 
$\uvc{l}$ 
can be conveniently expressed
in terms of Euler angles as follows
\begin{subequations}
  \label{eq:director}
\begin{align}
&
  \label{eq:director-d}
  \uvc{d}=\sin\theta_{\dd}\cos\phi_{\dd}\,\uvc{x}
+\sin\theta_{\dd}\sin\phi_{\dd}\,\uvc{y}
+\cos\theta_{\dd}\,\uvc{z},
\\
&
  \label{eq:director-m}
  \uvc{m}=\cos\gamma_{\dd}\,\vc{e}_{x}(\uvc{d})
+\sin\gamma_{\dd}\,\vc{e}_{y}(\uvc{d}),
\\
&
\label{eq:director-l}
  \uvc{l}=-\sin\gamma_{\dd}\,\vc{e}_{x}(\uvc{d})+
\cos\gamma_{\dd}\,\vc{e}_{y}(\uvc{d}),
\end{align}
\end{subequations}
where
$ \vc{e}_{x}(\uvc{d})=(\cos\theta_{\dd}\cos\phi_{\dd},
\cos\theta_{\dd}\sin\phi_{\dd},
-\sin\theta_{\dd})$
and
$ \vc{e}_{y}(\uvc{d})=(-\sin\phi_{\dd}, \cos\phi_{\dd}, 0)$.

We shall need to write the Maxwell equations for a harmonic
electromagnetic wave (time-dependent factor is $\exp\{-i\omega t\}$)
in the form:
\begin{subequations}
  \label{eq:maxwell}
\begin{align}
&
  \label{eq:maxwell-1}
   \bs{\nabla}\times\vc{E}=i \mu k_{\vac} \vc{H},
\\
&
\label{eq:maxwell-2}
  \bs{\nabla}\times\vc{H}=-i k_{\vac}  \vc{D},
\end{align}
\end{subequations}
where 
$k_{\vac}=\omega/c$ is the free-space wave number;
$\mu$ is the magnetic permittivity and
$\vc{D}=\bs{\varepsilon}\cdot\vc{E}$ is the electric displacement field.
The medium surrounding the NLC cell is assumed to be optically
isotropic and characterized by the dielectric constant $\epsilon_{\med}$
and the magnetic permittivity $\mu_{\med}$.
So, Maxwell's equations in the medium outside the cell 
can be obtained from Eq.~\eqref{eq:maxwell} by 
replacing $\mu$ and $\vc{D}$ with $\mu_{\med}$ and
$\epsilon_{\med} \vc{E}$, respectively. 

\begin{figure*}[!tbh]
%\vskip5mm
\centering
\resizebox{130mm}{!}{\includegraphics*{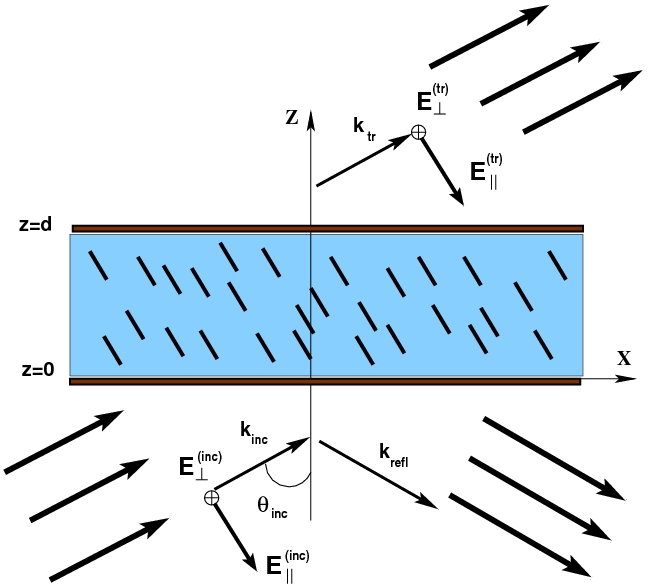}}
\caption{%
Geometry of nematic cell in the plane of incidence.
}
\label{fig:geom}
\end{figure*}

Referring to Fig.~\ref{fig:geom}, 
there are two plane waves in the lower half space
$z\le 0$ bounded by the entrance face of the NLC cell: 
the \textit{incoming incident wave} $\{\vc{E}_{\inc}, \vc{H}_{\inc}\}$
and the \textit{outgoing reflected wave} $\{\vc{E}_{\refl}, \vc{H}_{\refl}\}$.
The \textit{transmitted plane wave}  $\{\vc{E}_{\trans}, \vc{H}_{\trans}\}$ 
is excited by the incident wave and propagates along
the direction of incidence in the upper half space $z\ge d$ after the exit face. 
So, the electric field outside the cell is 
a superposition of the plane wave solutions
of the Maxwell equations 
\begin{subequations}
  \label{eq:E-med}
\begin{align}
&
  \label{eq:E-before}
  \vc{E}\vert_{z<0}=
\vc{E}_{\inc}(\uvc{k}_{\inc})
\ee^{i(\vc{k}_{\inc}\cdot\vc{r})}
+
\vc{E}_{\refl}(\uvc{k}_{\refl})
\ee^{i(\vc{k}_{\refl}\cdot\vc{r})},
\\
&
 \label{eq:E-after}
  \vc{E}\vert_{z>d}=
\vc{E}_{\trans}(\uvc{k}_{\trans})
\ee^{i(\vc{k}_{\trans}\cdot\vc{r})},
\end{align}
\end{subequations}
where the wave vectors
$\vc{k}_{\inc}$, $\vc{k}_{\refl}$ and $\vc{k}_{\trans}$
that are constrained to lie in the plane of incidence 
due to the boundary conditions which require 
the tangential components of the electric and magnetic
fields to be continuous at the boundary (interface) surfaces.
These conditions are given by 
\begin{subequations}
  \label{eq:bc-gen}
\begin{align}
&
  \label{eq:bc-E}
  \mvc{P}(\uvc{z})\cdot\bigl[
\vc{E}\vert_{z=0+0}-\vc{E}\vert_{z=0-0}
\bigr]
=  \mvc{P}(\uvc{z})\cdot\bigl[
\vc{E}\vert_{z=d+0}-\vc{E}\vert_{z=d-0}
\bigr]=0,
\\
&
 \label{eq:bc-H}
  \mvc{P}(\uvc{z})\cdot\bigl[
\vc{H}\vert_{z=0+0}-\vc{H}\vert_{z=0-0}
\bigr]
=  \mvc{P}(\uvc{z})\cdot\bigl[
\vc{H}\vert_{z=d+0}-\vc{H}\vert_{z=d-0}
\bigr]=0,
\end{align}
\end{subequations}
where
$\mvc{P}(\uvc{z})=\mvc{I}_3-\uvc{z}\otimes\uvc{z}$
is the projector onto 
the plane with the normal directed along the vector $\uvc{z}$
(the $x$-$y$ plane). 

Another consequence of the boundary conditions~\eqref{eq:bc-gen}
is that the tangential components of the wave vectors are the same.
Assuming that the incidence plane is the $x$-$z$
plane we have
\begin{align}
  \label{eq:k-alp}
\vc{k}_{\alpha}=k_{\med}\uvc{k}_{\alpha}=
k_x\,\uvc{x}+
k_z^{(\alpha)}\,\uvc{z},
\quad
\alpha\in\{\inc, \refl, \trans\},   
\end{align}
where $k_{\med}/k_{\vac}=n_{\med}=\sqrt{\mu_{\med}\epsilon_{\med}}$
is the refractive index of the ambient medium
and the components can be expressed
in terms of the incidence angle
$\theta_{\inc}$ as follows
\begin{align}
&
  \label{eq:k_x}
  k_x=k_{\med}\sin\theta_{\inc},
\\
&
\label{eq:k_z-alp}
k_z^{(\inc)}=k_z^{(\trans)}=-k_z^{(\refl)}=k_{\med}\cos\theta_{\inc}.
\end{align}
The plane wave 
traveling in the isotropic ambient medium
along the wave vector~\eqref{eq:k-alp} is transverse, so that
the polarization vector is given by 
\begin{align}
&
  \label{eq:E-alp}
  \vc{E}_{\alpha}(\uvc{k}_{\alpha})=
  E_{\parallel}^{(\alpha)}\vc{e}_x(\uvc{k}_{\alpha})+
  E_{\perp}^{(\alpha)}\vc{e}_y(\uvc{k}_{\alpha}),
\\
&
\label{eq:e_x-alp}
\vc{e}_x(\uvc{k}_{\alpha})=
k_{\med}^{-1}
\bigl(
k_z^{(\alpha)}\,\uvc{x}-k_x\,\uvc{z}
\bigr),
\quad
\vc{e}_y(\uvc{k}_{\alpha})=\uvc{y},
\end{align}
where $E_{\parallel}^{(\alpha)}$ and
$E_{\perp}^{(\alpha)}$ are the in-plane and out-of-plane
components of the electric field, respectively.
The vector characterizing the magnetic field is
\begin{align}
\label{eq:H-alp}
  \mu_{\med}\, \vc{H}_{\alpha}(\uvc{k}_{\alpha})=
\vc{q}_{\alpha}\times\vc{E}_{\alpha}(\uvc{k}_{\alpha})=
n_{\med}\,
\bigl[
  E_{\parallel}^{(\alpha)} \uvc{y}-
  E_{\perp}^{(\alpha)} \vc{e}_x(\uvc{k}_{\alpha})
\bigr],
\end{align}
where $\vc{q}_{\alpha}=k_{\vac}^{-1} \vc{k}_{\alpha}=
n_{\med}\uvc{k}_{\alpha}$.

%%%%%%%%%%%%%%%%%%%%%%%%%%%%%%%%%%%
\subsection{Operator of evolution and transmission matrix}
\label{subsec:transm-matrix}
%%%%%%%%%%%%%%%%%%%%%%%%%%%%%%%%%%%

The expressions~\eqref{eq:E-med}--\eqref{eq:H-alp}
give the electromagnetic field of incident,
transmitted and reflected waves propagating in the ambient medium.
This field is of the general form 
\begin{align}
  \label{eq:EH-form}
  \vc{E}(x,z)=\vc{E}(z)\,\ee^{i k_x x},
\quad
  \vc{H}(x,z)=\vc{H}(z)\,\ee^{i k_x x}
\end{align}
that can also be used for 
the field inside the NLC cell when the dielectric
tensor~\eqref{eq:diel-tens}
is independent of the in-plane coordinates $x$ and $y$.

The boundary conditions~\eqref{eq:bc-gen}
can be conveniently written using matrix notations as follows
\begin{subequations}
  \label{eq:bc-matr}
\begin{align}
&
  \label{eq:bc-matr-in}
  \vc{F}_{<}=\mvc{Q}_{\med}
\begin{pmatrix}
\vc{E}_{\inc}\\ \vc{E}_{\refl}
\end{pmatrix}
=\vc{F}(0),
\\
&
\label{eq:bc-matr-out}
\vc{F}_{>}=\mvc{Q}_{\med}
\begin{pmatrix}
\vc{E}_{\trans}\\ \vc{0}
\end{pmatrix}
=\vc{F}(d),
\end{align}
\end{subequations}
where
\begin{align}
  \label{eq:matr-notat}
  \vc{F}(z)\equiv \begin{pmatrix}
E_x(z)\\ E_y(z)\\H_y(z)\\-H_x(z)
\end{pmatrix},
\quad
\vc{E}_{\alpha}\equiv\begin{pmatrix}
E_{\parallel}^{(\alpha)}\\ E_{\perp}^{(\alpha)}
\end{pmatrix}.
\end{align}
Our notations can be regarded as a version of 
the well known $4\times 4$ matrix formalism 
due to Berreman~\cite{Berreman:josa:1972}.

The matrix $\mvc{Q}_{\med}$ 
relates the field vectors,
$\vc{F}_{<}\equiv\vc{F}(0-0)$ and $\vc{F}_{>}\equiv\vc{F}(d+0)$,
and the vector amplitudes $\vc{E}_{\alpha}$
of the waves in the surrounding medium.
Its block structure 
\begin{align}
  \label{eq:Qm-block}
  \mvc{Q}_{\med}=
\begin{pmatrix}
\mvc{E}_{+}^{(m)} & \mvc{E}_{-}^{(m)}\\
\mvc{H}_{+}^{(m)} & \mvc{H}_{-}^{(m)}\\
\end{pmatrix}
\end{align}
is characterized by the four $2\times 2$ matrices:
$\mvc{E}_{\pm}^{(m)}$ and $\mvc{H}_{\pm}^{(m)}$.
From Eqs.~\eqref{eq:E-alp}-\eqref{eq:H-alp}
it can be readily obtained that these matrices  
are diagonal and are given by
\begin{subequations}
\label{eq:HE_med}
\begin{align}
&
\label{eq:H_m}
  \mvc{H}_{+}^{(m)}=\bs{\sigma}_3\,\mvc{H}_{-}^{(m)}=\mu_{\med}^{-1} n_{\med}\,\mvc{A},
\\
&
  \label{eq:E_m}
  \mvc{E}_{+}^{(m)}=-\bs{\sigma}_3\,\mvc{E}_{-}^{(m)}=\mu_{\med}^{-1} n_{\med}\,\mvc{B},
\end{align}
\end{subequations}
where $\bs{\sigma}_3=\diag(1,-1)$,
$\mvc{A}=\diag(1,\cos\theta_{\inc})$
and $\mvc{B}=\mu_{\med} n_{\med}^{-1} \diag(\cos\theta_{\inc},1)$.
It is rather straightforward to check the validity of the
orthogonality relations for the
matrices~\eqref{eq:HE_med}
\begin{align}
  \label{eq:orthog-rel}
  \hcnj{\bigl[\mvc{E}_{\mu}^{(\med)}\bigr]}\mvc{H}_{\nu}^{(\med)}
+
  \hcnj{\bigl[\mvc{H}_{\mu}^{(\med)}\bigr]}\mvc{E}_{\nu}^{(\med)}
=\delta_{\mu \nu}\,\mvc{N}_{\mu}^{(m)},
\quad
\mu, \nu \in\{+,-\},
\end{align}
where $\mvc{N}_{\pm}^{(m)}=\pm 2\mu_{\med}^{-1} n_{\med}\cos\theta_{\inc}\,\mvc{I}_2$, 
$\delta_{\mu\nu}$ is the Kronecker symbol and
the superscript $\dagger$ indicates Hermitian conjugation.

As it was shown in Ref.~\cite{Oldano:pra:1989},
the relations~\eqref{eq:orthog-rel}, 
where $\mvc{N}_{\med}$ is a diagonal matrix,
can be deduced
from the conservation law for the energy flux in non-absorbing media
that, in the case of one-dimensional inhomogeneity, 
requires the $z$ component of the Poynting vector to be independent of $z$.
The relations~\eqref{eq:orthog-rel} can be rewritten in the more
compact matrix form:
\begin{align}
  \label{eq:orth-matr}
  \hcnj{[\mvc{Q}_{\med}]}\mvc{G}\mvc{Q}_{\med}=\mvc{N}_{\med},
\quad
\mvc{G}\equiv\begin{pmatrix}\vc{0} & \mvc{I}_2\\
\mvc{I}_2 & \vc{0}
\end{pmatrix},
\end{align}
where 
$\mvc{N}_{\med}=\diag(\mvc{N}_{+}^{(m)},\mvc{N}_{-}^{(m)})=N_{\med}\diag(\mvc{I}_2,-\mvc{I}_2)$
and $N_{\med}=2\mu_{\med}^{-1} n_{\med}\cos\theta_{\inc}$.
Algebraically, Eq.~\eqref{eq:orth-matr} can be used to simplify
 inversion of the matrix $\mvc{Q}_{\med}$
 \begin{equation}
   \label{eq:Qm-inv}
 \mvc{Q}_{\med}^{-1}=(2\cos\theta_{\inc})^{-1}\diag(\mvc{I}_2,\bs{\sigma}_3)
\begin{pmatrix}
\mvc{A} & \mvc{B}\\
-\mvc{A} & \mvc{B}
\end{pmatrix}  
 \end{equation}
and to ease qualitative analysis  
[see , e.g., Ref.~\cite{Yakovl:opsp:1998}
for a more extended discussion of applications].

Given the characteristics of the incident wave,
the amplitudes of the reflected and transmitted waves
cannot be computed from the boundary conditions~\eqref{eq:bc-matr}
unless the electromagnetic field within the NLC cell is defined.
At this stage we just note that,
for the in-plane
components of the electric and magnetic field,
$\{E_x,E_y\}$ and $\{H_x,H_y\}$,
the result can be presented in the form of a general solution
to the linear problem
\begin{align}
  \label{eq:oper-evol}
  \vc{F}(z)=\mvc{U}(z)\vc{F}(0),
\end{align}
where the matrix-valued function $\mvc{U}(z)$ 
is the \textit{evolution operator} 
which can be computed by solving the Maxwell
equations~\eqref{eq:maxwell}.

After substituting Eq.~\eqref{eq:oper-evol} into the
boundary conditions~\eqref{eq:bc-matr-in} and using the orthogonality
relations~\eqref{eq:orthog-rel}, we have
\begin{align}
&
  \label{eq:rel-oper}
  \begin{pmatrix}
\vc{E}_{\inc}\\ \vc{E}_{\refl}
\end{pmatrix}
=\mvc{V}
\begin{pmatrix}
\vc{E}_{\trans}\\ \vc{0}
\end{pmatrix},
\\
&
  \label{eq:V-op}
  \mvc{V}=\mvc{N}_{\med}^{-1}\hcnj{[\mvc{Q}_{\med}]}\mvc{G}\mvc{U}^{-1}
\mvc{Q}_{\med}=
\begin{pmatrix}
\mvc{V}_{ee} & \mvc{V}_{eh}\\ 
\mvc{V}_{he} & \mvc{V}_{hh}
\end{pmatrix},
\end{align}
where $\mvc{U}^{-1}\equiv \mvc{U}^{-1}(d)$ and
$\mvc{Q}_{\med}=\mu_{\med}^{-1} n_{\med}
\begin{pmatrix}
\mvc{B} & -\mvc{B}\\ 
\mvc{A} & \mvc{A}
\end{pmatrix}
\diag(\mvc{I}_{2},\bs{\sigma}_3)
$.

From Eq.~\eqref{eq:rel-oper} we have
\begin{align}
  \label{eq:transm-rel}
  \begin{pmatrix}
E_{\parallel}^{(\trans)}\\
E_{\perp}^{(\trans)}
\end{pmatrix}
&
=\mvc{T}
\begin{pmatrix}
E_{\parallel}^{(\inc)}\\
E_{\perp}^{(\inc)}
\end{pmatrix},
\\
  \label{eq:trans-mat1}
 \mvc{T}
&
=\mvc{V}_{ee}^{-1},
\end{align}
where $\mvc{T}$ is the \textit{transmission (transfer) matrix}
linking the transmitted and incident waves.

Similar result for the reflected wave is
\begin{align}
&
  \label{eq:refl-rel}
  \begin{pmatrix}
E_{\parallel}^{(\refl)}\\
E_{\perp}^{(\refl)}
\end{pmatrix}
=
\mvc{R}
\begin{pmatrix}
E_{\parallel}^{(\inc)}\\
E_{\perp}^{(\inc)}
\end{pmatrix},
\\
&
  \label{eq:refl-mat1}
 \mvc{R}=\mvc{V}_{he}\mvc{V}_{ee}^{-1}=\mvc{V}_{he}\mvc{T},
\end{align}
where $\mvc{R}$ is the \textit{reflection matrix}.

%%%%%%%%%%%%%%%%%%
\subsection{Eigenmodes}
\label{subsec:obliq-incid-plane}
%%%%%%%%%%%%%%%%%%

The equations~\eqref{eq:trans-mat1} and~\eqref{eq:refl-mat1}
give the transmission 
and reflection matrices expressed in terms of the matrix~\eqref{eq:V-op}
which is related to  the evolution operator~\eqref{eq:oper-evol}.
For our purposes, it is instructive to see how
this operator can be derived by using the basis of 
\textit{eigenmodes (normal modes)}
in a uniformly anisotropic medium.
In this case the eigenmodes are known to be
linearly polarized plane waves characterized by 
the $\uvc{k}$ dependent 
refractive indices~\cite{Born,Yariv:bk:1984,Azz:1977}. 

For plane waves, owing to the Maxwell equation~\eqref{eq:maxwell-2},
the electric displacement field is transverse,
$\sca{\vc{D}}{\vc{k}}=0$. So, assuming that the wave
vector $\vc{k}$ lies in the plane of incidence (the $x$-$z$ plane), 
the vector $\vc{D}$ can be conveniently defined by its components
in the basis $\{e_{x}(\uvc{k}), e_{y}(\uvc{k}), e_{z}(\uvc{k})\equiv\uvc{k}\}$
\begin{align}
&
  \label{eq:D-transv}
  \vc{D}=D_{x} e_{x}(\uvc{k}) + D_{y} e_{y}(\uvc{k}),
\\
&
\label{eq:kxz}
\uvc{k}\equiv e_{z}(\uvc{k})=
k^{-1} \vc{k}=k^{-1} (k_{x}\uvc{x}+k_{z}\uvc{z})
= q^{-1}\vc{q},
\end{align}
where 
$e_x(\uvc{k})=q^{-1} (q_{z}\uvc{x}-q_{x}\uvc{z})$,
$e_y(\uvc{k})=\uvc{y}$, $q_{x, z}=k_{x, z}/k_{\vac}$ and 
$q=k/k_{\vac}=n$ is the refractive index.
The vector $\vc{q}$ defined in Eq.~\eqref{eq:kxz}
is parallel to $\vc{k}$ and its length is equal to the refractive index.
For convenience, we shall often use this vector
in place of the wave vector.

Given the electric displacement field~\eqref{eq:D-transv},
the electric field can be found from the constitutive relation
\begin{align}
&
  \label{eq:D-E}
  \vc{E}=\mu\,\bs{\eta}\cdot\vc{D},
\\
&
\label{eq:tens-eta}
  \bs{\eta}=\eta_3\,\mvc{I}_3+\Delta\eta_1\,\uvc{d}\otimes\uvc{d}
+\Delta\eta_2\,\uvc{m}\otimes\uvc{m},
\end{align}
where $\eta_i=(\mu\epsilon_i)^{-1}$,
$\Delta\eta_i=\eta_i-\eta_3$ and $\mu\bs{\eta}$ is the inverse
dielectric tensor ($\mu\,\bs{\eta}\cdot\bs{\varepsilon}=\mvc{I}_3$).

The Maxwell equations can now be combined with the
relation~\eqref{eq:D-E}
to yield the equation for the displacement vector $\vc{D}$
in the form of an eigenvalue problem 
\begin{align}
  \label{eq:eig-eq}
  \bs{\eta}_{t}\cdot\vc{D}=q^{-2}\vc{D},
\quad
  \bs{\eta}_{t}=\mvc{P}(\uvc{k})\cdot
  \bs{\eta}\cdot\mvc{P}(\uvc{k}),
\end{align}
where
$\mvc{P}(\uvc{k})=\mvc{I}_3-\uvc{k}\otimes\uvc{k}$
is the projector onto 
the plane normal to the wave vector $\vc{k}$.
By using the expression for the inverse dielectric
tensor~\eqref{eq:tens-eta}
we can rewrite the equation~\eqref{eq:eig-eq}
in the explicit matrix form:
\begin{align}
&
  \label{eq:matr-eigeq}
  \bigl[
a_{-}\,\bs{\sigma}_3+b\,\bs{\sigma}_1-\lambda\mvc{I}_2
\bigr]
\begin{pmatrix}
  D_{x}\\D_{y}
\end{pmatrix}
=\vc{0},
\quad
\bs{\sigma}_1=
\begin{pmatrix}
  0 & 1\\1 &0
\end{pmatrix},
\\
&
\label{eq:coef-a}
2 a_{\mp} = \Delta\eta_1(\tilde{d}_x^2\mp\tilde{d}_y^2)+
\Delta\eta_2(\tilde{m}_x^2\mp\tilde{m}_y^2),
\\
&
\label{eq:coef-b}
b = \Delta\eta_1\tilde{d}_x\tilde{d}_y
+\Delta\eta_2\tilde{m}_x\tilde{m}_y,
\\
&
\label{eq:coef-lmb}
\lambda=1-\eta_3 q^2-a_{+},
\end{align}
where $\tilde{d}_{x, y}=q\sca{\uvc{d}}{e_{x, y}(\uvc{k})}$
 and $\tilde{m}_{x, y}=q\sca{\uvc{m}}{e_{x, y}(\uvc{k})}$.
Then the dispersion relation (Fresnel's equation)
\begin{align}
   \label{eq:eigval-qz}
   (1-\eta_3 q^2)
\bigl[
1-\eta_3 q^2 -\Delta\eta_1(q^2-\tilde{d}_z^2) -
\notag
\\
\Delta\eta_2(q^2-\tilde{m}_z^2)
\bigr]+\Delta\eta_1\Delta\eta_2
q^2 \tilde{l}_z^2=0,
 \end{align}
where $q \tilde{l}_z= \tilde{d}_x\tilde{m}_y   -
\tilde{d}_y\tilde{m}_x=q^2\sca{\uvc{l}}{\uvc{k}}$,
can be derived as the condition for 
the system of linear equations~\eqref{eq:matr-eigeq}
to have a non-vanishing solution.

The Fresnel equation describes the wave surface.
In our case solving the algebraic equation~\eqref{eq:eigval-qz}
at $q_x=n_{\med}\sin\theta_{\inc}$
gives the values of the $z$ component of the vector $\vc{q}$, $q_z$.

Generally, there are four roots of Eq.~\eqref{eq:eigval-qz}
$q_z^{(\alpha)}$.
Each root corresponds to the eigenwave propagating 
inside the cell with the wave vector $\vc{k}_{\alpha}=
k_{\vac}(q_x,0,q_z^{(\alpha)})$ and the refractive index
$n_{\alpha}=q_{\alpha}$.
The corresponding polarization vector of the electric displacement
field is given by
\begin{align}
  \label{eq:D_i}
  \vc{D}^{(\alpha)}=
\begin{cases}
\cos\phi_{\alpha}\, e_{x}(\uvc{k}_{\alpha}) + 
\sin\phi_{\alpha}\, e_{y}(\uvc{k}_{\alpha}),
& \lambda_{\alpha}>0,
\\
-\sin\phi_{\alpha}\, e_{x}(\uvc{k}_{\alpha}) + 
\cos\phi_{\alpha}\, e_{y}(\uvc{k}_{\alpha}),
& \lambda_{\alpha}<0,
\end{cases}
\end{align}
where
\begin{align}
  \label{eq:phi_i}
\lambda_{\alpha}=\lambda\vert_{q_z=q_z^{(\alpha)}},
\quad
  2\phi_{\alpha}=\arg(a_{-}+i b)\vert_{q_z=q_z^{(\alpha)}}.
\end{align}
From Eq.~\eqref{eq:phi_i} it is clear that the azimuthal angle
$\phi_{\alpha}$ becomes indeterminate in the degenerate case
when the coefficients $a_{-}$ and $b$ are both identically 
equal to zero. Typically, as far as the eigenmodes are concerned, 
this case does not present any fundamental difficulties.
It just means that the azimuthal angles of the degenerate eigenmodes
can be prescribed arbitrarily. Such freedom of choice, however, 
does not affect the evolution operator which remains uniquely defined.

The procedure to determine the characteristics
of the eigenmodes involves the following steps:
(a)~evaluation of $q_z$ by solving the Fresnel
equation~\eqref{eq:eigval-qz};
(b)~calculation of the polarization vectors of the electric
displacement field $\vc{D}^{(\alpha)}$ by 
using the formula~\eqref{eq:D_i};
(c)~computing the polarization vectors of the electric and magnetic fields
from the relations:
$\vc{E}^{(\alpha)}=\mu\,\bs{\eta}\cdot\vc{D}^{(\alpha)}$
(see Eq.~\eqref{eq:D-E}) and 
$\mu\,\vc{H}^{(\alpha)}=\vc{q}_{\alpha}\times\vc{E}^{(\alpha)}$
(see Eq.~\eqref{eq:H-alp}).

%%%%%%%%%%%%%%%%%%%%%%%%%%%%%%%%
\subsubsection{Uniaxial anisotropy with $\Delta\epsilon_2=0$}
\label{subsubsec:uniax-anis}
%%%%%%%%%%%%%%%%%%%%%%%%%%%%%%%%

Now we apply the above procedure to the limiting case
of uniaxial anisotropy with $\epsilon_2=\epsilon_3$
and $\Delta\epsilon_2=\Delta\eta_2=0$.
At $\Delta\eta_2=0$,
the Fresnel equation~\eqref{eq:eigval-qz}
takes the factorized form and 
the values of $q_z$ can be found as roots of
two quadratic equations.

The first equation $1-\eta_3 q^2=0$ represents the spherical wave
surface. The corresponding eigenmodes are known as 
the \textit{ordinary waves}.
There are two  values of $q_z$  
\begin{align}
  \label{eq:qz_pmo}
  q_{z}^{(\pm o)}=\pm\sqrt{n_{\perp}^2-q_x^2},
\end{align}
where $n_{\perp}^2=\mu\epsilon_3\equiv\mu\epsilon_{\perp}$,
that are equal in value but opposite in sign.
When, similar to the incident and transmitted waves, 
the $z$ component of the wave vector (and the vector $\vc{q}$)
is positive, the eigenmode might be called the 
\textit{refracted (forward) eigenwave}.
In the opposite case where, similar to the reflected wave, 
$q_{z}^{(\alpha)}$ is negative, the eigenmode will be referred to as 
the \textit{reflected (backward)  eigenwave}.
So, Eq.~\eqref{eq:qz_pmo} describes two ordinary eigenmodes:
the refracted eigenwave with $q_z=q_{z}^{(+o)}>0$ and the reflected
eigenwave with $q_z=q_{z}^{(- o)}<0$.

The second equation 
\begin{align}
  \label{eq:qze-eq}
  q^2+u_a\sca{\vc{q}}{\uvc{d}}^2-n_{\parallel}^2=0,
\end{align}
where
$n_{\parallel}^2=\mu\epsilon_1\equiv\mu\epsilon_{\parallel}$
and 
$u_a=-\Delta\eta_1/\eta_1=(n_{\parallel}^2-n_{\perp}^2)/n_{\perp}^2$
is the \textit{anisotropy parameter},
gives the values of $q_z$ for the eigenmodes known as 
the \textit{extraordinary waves}.
These are given by
\begin{align}
& 
 \label{eq:qz_pme}
  q_{z}^{(\pm e)}=[1+u_a d_z^2]^{-1}
\bigl\{
-u_{a} d_z d_x q_x\pm\sqrt{D}
\bigr\},
\\
&
\label{eq:discrim}
D = n_{\parallel}^2 
\bigl(1+u_a d_z^2\bigr) -
q_x^2 
\bigl[1+u_a (d_x^2+d_z^2)\bigr] 
\end{align}
where $d_x=\sca{\uvc{d}}{\uvc{x}}$ and
$d_z=\sca{\uvc{d}}{\uvc{z}}$.

At $u_a>0$ ($u_a<0$), in the $x$-$z$ plane, Eq.~\eqref{eq:qze-eq}
describes the ellipse with the major (minor) semi-axis of the length
$n_{\parallel}$ oriented perpendicular to
the projection of the director~\eqref{eq:director-d} on the plane of incidence
$(d_x,0,d_z)$. The length of minor (major) semi-axis, 
$\tilde{n}_{\perp}=[n_{\perp}^{-2}-u_a (d_y/n_{\parallel})^2]^{-1/2}$,
depends on the $y$ component of the director
and varies from $n_{\perp}$ to $n_{\parallel}$
as $d_y^2$ increases from zero to unity.
Clearly, degeneracy in refractive indices with $n_o=n_{\pm e}$ 
may occur only if the director is in the incidence plane
($\phi_{\dd}=0$). Additionally,
the matching condition for the $x$ components
of $\vc{q}$ and the director 
$q_x\equiv n_m\sin\theta_{\inc}=\pm n_{o} d_x\equiv
\pm n_o\sin\theta_{\dd}$ needs to be met.

The wave vectors and the refractive indices
of the normal modes are determined by the relation
\begin{equation}
  \label{eq:q_pm}
  \vc{q}_{\pm\alpha}= k_{\vac}^{-1}\vc{k}_{\pm\alpha}=
q_x\,\uvc{x}+q_{z}^{(\pm\alpha)}\,\uvc{z}=
n_{\pm\alpha}\uvc{k}_{\pm\alpha},
\quad \alpha\in\{o, e\},
\end{equation}
where $n_{\pm\alpha}=q_{\pm\alpha}$,
$n_{\pm o}=n_o=n_{\perp}$ is the ordinary refractive index
and
$n_{\pm e}$ is the refractive index of the extraordinary wave propagating
along the unit vector $\uvc{k}_{\pm e}$.

The relations
\begin{align}
&
  \label{eq:sing-lamb}
  \sign\lambda_{\pm o}=-\sign\Delta\eta_1,
\quad
  \sign\lambda_{\pm e}=\sign\Delta\eta_1,
\\
&
\label{eq:a_ib}
a_{-}+i b=\Delta\eta_1 (\tilde{d}_x+i\tilde{d}_y)^2
\end{align}
obtained by substituting $\Delta\eta_2=0$ into
Eqs.~\eqref{eq:coef-a}-\eqref{eq:coef-lmb}
can now be combined with Eq.~\eqref{eq:D_i}
to yield the following result for the polarization vectors of
the electric displacement field: 
\begin{align}
&
  \label{eq:eigv-o}
  \vc{D}^{(\pm o)}\propto 
-\tilde{d}_y e_{x}(\uvc{k}_{\pm e}) + \tilde{d}_x
e_{y}(\uvc{k}_{\pm e})
\propto \uvc{k}_{\pm o}\times\uvc{d},
\\
&
  \label{eq:eigv-e}
  \vc{D}^{(\pm e)}\propto 
\tilde{d}_x e_{x}(\uvc{k}_{\pm e}) + \tilde{d}_y
e_{y}(\uvc{k}_{\pm e})
\propto
\mvc{P}(\uvc{k}_{\pm e})\cdot\uvc{d}.
\end{align}
Following the procedure described at the end of
Sec.~\ref{subsec:obliq-incid-plane}
we then find the polarization vectors of the electric field
for the eigenmodes 
\begin{subequations}
\label{eq:E_eig-vect}
\begin{align}
&
  \label{eq:E-ordn}
  \vc{E}^{(\pm o)}=-\vc{q}_{\pm o}\times\uvc{d},
\\
&
  \label{eq:E-extr}
  \vc{E}^{(\pm e)}=
\bigl[
\uvc{d} -
n_o^{-2}
\sca{\uvc{d}}{\vc{q}_{\pm e}}\vc{q}_{\pm e}
\bigr],
\end{align}
\end{subequations}
where Eq.~\eqref{eq:E-extr} was derived by using the identity
\begin{equation}
  \label{eq:aux-rel1}
  \bs{\eta}\cdot\mvc{P}(\uvc{k})\uvc{d}=n^{-2}\uvc{d}-n_{\perp}^{-2}
\sca{\uvc{d}}{\uvc{k}}\uvc{k}
\end{equation}
with $n^{-2}\equiv q^{-2}=\bigl[1+u_{a}\sca{\uvc{d}}{\uvc{k}}^2\bigr]
n_{\parallel}^{-2}$.

The result for the magnetic field of the normal modes is
\begin{subequations}
\label{eq:H_eig-vect}
\begin{align}
&
\label{eq:H-ordn}
\vc{H}^{(\pm o)}=\mu^{-1}
\bigl[n_o^{2}\,\uvc{d}- 
\sca{\uvc{d}}{\vc{q}_{\pm o}}\vc{q}_{\pm o}
\bigr],
\\
&
\label{eq:H-extr}
  \vc{H}^{(\pm e)}=
\mu^{-1}\, 
\vc{q}_{\pm e}\times\uvc{d}.
\end{align}
\end{subequations}

In order to present the results in the matrix form, we,
by analogy with Eq.~\eqref{eq:Qm-block}, define the matrix
composed of the polarization vectors as follows
\begin{align}
  \label{eq:Q-nlc}
  \mvc{Q}=
\begin{pmatrix}
\mvc{E}_{+} & \mvc{E}_{-}\\
  \mvc{H}_{+}&  \mvc{H}_{-}
\end{pmatrix},
\end{align}
where
\begin{align}
  \label{eq:E_pm}
  \mvc{E}_{\pm}=
\begin{pmatrix}
E^{(\pm o)}_x & E^{(\pm e)}_x\\
E^{(\pm o)}_y & E^{(\pm e)}_y
\end{pmatrix},
\quad
  \mvc{H}_{\pm}=
\begin{pmatrix}
H^{(\pm o)}_y & H^{(\pm e)}_y\\
-H^{(\pm o)}_x & -H^{(\pm e)}_x
\end{pmatrix}.
\end{align}
The explicit expressions for
the in-plane components of the electric field are given by
\begin{subequations}
\label{eq:E_eig-xy}
\begin{align}
&
  \label{eq:E_o_xy}
  E_{x}^{(\pm o)} = q_z^{(\pm o)} d_y,
\quad
E_{y}^{(\pm o)}=-[\vc{q}_{\pm o}\times\uvc{d}]_{y},
\\
&
\label{eq:E_e_xy}
  E_{x}^{(\pm e)} =
n_o^{-2}
\bigl\{
 (n_{o}^{2}-n_{\pm e}^{2}) d_x+
q_z^{(\pm e)} [\vc{q}_{\pm e}\times\uvc{d}]_{y}
\bigr\},
\quad
E_{y}^{(\pm e)}=d_{y}.
\end{align}
\end{subequations}
Those for the magnetic field are
\begin{subequations}
\label{eq:H_eig-xy}
\begin{align}
&
  \label{eq:H_o_xy}
  \mu H_{x}^{(\pm o)} =q_z^{(\pm o)} [\vc{q}_{\pm o}\times\uvc{d}]_{y},
\quad
\mu H_{y}^{(\pm o)}= n_o^2 d_y,
\\
&
\label{eq:H_e_xy}
  \mu H_{x}^{(\pm e)} = - q_z^{(\pm e)} d_y,
\quad
\mu H_{y}^{(\pm e)}=[\vc{q}_{\pm e}\times\uvc{d}]_{y},
\end{align}
\end{subequations}
where $[\vc{q}_{\alpha}\times\uvc{d}]_{y}=q_z^{(\alpha)} d_x - q_x d_z$.

The matrix of eigenvectors~\eqref{eq:Q-nlc} along with 
the eigenvalues given in Eq.~\eqref{eq:qz_pmo} and
Eq.~\eqref{eq:qz_pme}
can now be used to describe
the $z$ dependence of the in-plane components of the electromagnetic
field inside the cell in terms of the eigenmodes amplitudes.
So, we have
\begin{align}
&
  \label{eq:F-nlc}
  \vc{F}(z)=\mvc{Q}\,\mvc{U}_d(z)
\begin{pmatrix}
\bs{\beta}_{+}\\
\bs{\beta}_{-}
\end{pmatrix},
\quad
\mvc{U}_d(z)=
\begin{pmatrix}
\mvc{U}_{d}^{(+)}(z) & \vc{0}\\
\vc{0} & \mvc{U}_{d}^{(-)}(z)
\end{pmatrix},
\\
&
\label{eq:U_pm}
\mvc{U}_{d}^{(\pm)}(z)=\diag\bigl(\exp[i k_{z}^{(\pm o)} z],
\exp[i k_{z}^{(\pm e)} z]\bigr),
\end{align}
where $\bs{\beta}_{\pm}\equiv\begin{pmatrix}
\beta_{\pm o}\\
\beta_{\pm e}
\end{pmatrix}$
are the columns representing the amplitudes of the eigenmodes.

%%%%%%%%%%%%%%%%%%%%%%%%%
\subsubsection{Transmission and reflection matrices}
\label{subsubsec:transm-refl-matr}
%%%%%%%%%%%%%%%%%%%%%%%%%

Now it is not difficult to derive
the evolution operator~\eqref{eq:oper-evol}
from the equation~\eqref{eq:F-nlc}. 
We can also use
the orthogonality relation~\eqref{eq:orth-matr}
for the matrix~\eqref{eq:Q-nlc} so as to
obtain the following result
\begin{align}
  \label{eq:oprt-evol-nlc}
  \mvc{U}(z)=
\mvc{Q}\,\mvc{U}_d(z)\,\mvc{Q}^{-1}
=\mvc{Q}\,\mvc{U}_d(z)\,\mvc{N}^{-1}
\hcnj{\mvc{Q}}\mvc{G},
\end{align}
where 
the matrix $\mvc{G}$ is defined in Eq.~\eqref{eq:orth-matr}.

Owing to the orthogonality relations~\eqref{eq:orthog-rel},
the matrix 
$\mvc{N}=\hcnj{\mvc{Q}}\mvc{G}\mvc{Q}=\diag(\mvc{N}_{+},\mvc{N}_{-})$
is diagonal and its non-vanishing elements
\begin{align}
  \label{eq:N_pm_el}
  \mvc{N}_{\pm}=
\begin{pmatrix}
N_{\pm o} & 0\\ 0& N_{\pm e}
\end{pmatrix},
\quad N_{\alpha} = 2 \sca{\uvc{z}}{\vc{E}^{(\alpha)}\times\vc{H}^{(\alpha)}}
\end{align}
are proportional to the normal components of the Poynting vector
of the eigenmodes. 
Note that the evolution operator is not unitary.
It is clear from the identity
\begin{align}
&
  \label{eq:proper-U}
  \mvc{U}^{-1}(z)=\cnj{\mvc{U}}(z)=\mvc{G} \hcnj{\mvc{U}}(z)\mvc{G},
\end{align}
where an asterisk indicates complex conjugation,
that immediately follows from the expression~\eqref{eq:oprt-evol-nlc}. 

The operator~\eqref{eq:oprt-evol-nlc} can be substituted into
the expression for the matrix $\mvc{V}$~\eqref{eq:V-op} 
to yield the transmission and reflection
matrices given by Eq.~\eqref{eq:trans-mat1} and
Eq.~\eqref{eq:refl-mat1}, respectively.
By using Eq.~\eqref{eq:proper-U},
we have the relation for the matrix $\mvc{V}$
\begin{align}
  \label{eq:proper-V}
  \mvc{V}^{-1}=\cnj{\mvc{V}}=\mvc{G}_3 \hcnj{\mvc{V}}\mvc{G}_3,
\quad
\mvc{G}_3=\diag(\mvc{I}_2,-\mvc{I}_2).
\end{align}
The conservation law 
$\hcnj{\mvc{T}}\mvc{T}+\hcnj{\mvc{R}}\mvc{R}=\mvc{I}_2$
is a consequence of the identity~\eqref{eq:proper-V}
and we also arrive at the conclusion that the transmission
matrix~\eqref{eq:trans-mat1} is symmetric.

The explicit formulas for the transmission and reflection matrices
given by Eq.~\eqref{eq:trans-mat1} and Eq.~\eqref{eq:refl-mat1}.
can be obtained by substituting Eq.~\eqref{eq:oprt-evol-nlc}
into Eq.~\eqref{eq:V-op}.
After some straightforward algebraic manipulations
we have 
\begin{align}
&
  \label{eq:Tmatr-expl}
  \mvc{T}=2\mu_{\med}\,n_{\med}^{-1}\cos\theta_{\inc}\,
\bs{\tau}^{-1},
\quad
\bs{\tau}=\bs{\tau}_{+}+\bs{\tau}_{-},
\\
&
  \label{eq:Rmatr-expl}
  \mvc{R}=\bs{\sigma}_3
\Bigl[
\mvc{B}_{+}\tilde{\mvc{U}}_{+}\hcnj{[\mvc{A}_{+}]}
+\mvc{B}_{-}\tilde{\mvc{U}}_{-}\hcnj{[\mvc{A}_{-}]}
\Bigr] 
\bs{\tau}^{-1},
\end{align}
where
\begin{align}
&
\label{eq:tau_pm-tldU_pm}
\bs{\tau}_{\pm}=
\mvc{A}_{\pm}\tilde{\mvc{U}}_{\pm}\hcnj{[\mvc{A}_{\pm}]},
\quad
\tilde{\mvc{U}}_{\pm}=\cnj{[\mvc{U}_{d}^{(\pm)}(d)]} \mvc{N}_{\pm}^{-1},
\\
&
  \label{eq:AB_pm}
  \mvc{A}_{\pm}=\mvc{B}\mvc{H}_{\pm}+\mvc{A}\mvc{E}_{\pm},
\quad
  \mvc{B}_{\pm}=\mvc{B}\mvc{H}_{\pm}-\mvc{A}\mvc{E}_{\pm}.
\end{align}
Substituting the expressions for the electric and magnetic
fields of the eigenmodes 
[see Eq.~\eqref{eq:E_eig-vect} 
and Eq.~\eqref{eq:H_eig-vect}, respectively]
into Eq.~\eqref{eq:N_pm_el} gives
the diagonal elements of the matrices 
$\mvc{N}_{+}$ and $\mvc{N}_{-}$
\begin{align}
&
  \label{eq:N_pmo}
  \mu\,N_{\pm o}=
2 q_z^{(\pm o)}
\bigl[
n_o^2 - \sca{\vc{q}_{\pm o}}{\uvc{d}}^2
\bigr],
\\
&
  \label{eq:N_pme}
  \mu\,N_{\pm e}=
2 n_o^{-2}
\Bigl[
d_z \sca{\vc{q}_{\pm e}}{\uvc{d}}
(n_{\pm e}^2 - n_o^2)+
q_z^{(\pm e)}
\bigl(n_o^2 -
\sca{\vc{q}_{\pm e}}{\uvc{d}}^2
\bigr)
\Bigr].
\end{align}

The expressions~\eqref{eq:Tmatr-expl}--\eqref{eq:N_pme}
can now be used to write explicitly the elements of the transmission
matrix. The matrices $\mvc{E}_{\pm}$ and $\mvc{H}_{\pm}$
are given by Eqs.~\eqref{eq:E_pm}--\eqref{eq:H_eig-xy}
and their diagonal (non-diagonal) elements are zero
provided the director is normal (parallel) to the plane of incidence.
In these cases, where either $d_y=1$ or $d_y=0$,
the matrix $\bs{\tau}$ and the transmission matrix
are both diagonal
\begin{align}
  \label{eq:giag-transm}
  \bs{\tau}=\diag(\tau_x,\tau_y),
\quad
  \mvc{T}=\diag(t_x,t_y),
\end{align}
where 
$t_{x,\,y}=2\mu_{\med}\,n_{\med}^{-1}\cos\theta_{\inc}\tau_{x,\, y}^{-1}$.

The simplest case occurs when 
the cell is homeotropically aligned and $d_z=1$.
For the homeotropic director structure,
the elements of the matrix $\bs{\tau}$ are given by
\begin{subequations}
  \label{eq:tau-homeotr}
\begin{align}
  \label{eq:tau-x-homeo}
  \tau_x=&2\mu_{\med}\,n_{\med}^{-1}\cos\theta_{\inc}\cos\delta_{e}
-i [q_{z}^{(e)}]^{-1}\sin\delta_{e}\times
\notag
\\
&
\bigl[
(\mu_{\med} n_o n_{\med}^{-1}
\cos\theta_{\inc})^2+
(n_o^{-1}q_{z}^{(e)})^2
\bigr],
\\
  \label{eq:tau-y-homeo}
  \tau_y=&2\mu_{\med}\,n_{\med}^{-1}\cos\theta_{\inc}\cos\delta_{o}
-i [q_{z}^{(o)}]^{-1}\sin\delta_{o}\times
\notag
\\
&
\bigl[
\cos^2\theta_{\inc}+
(\mu_{\med} n_{\med}^{-1}q_{z}^{(o)})^2
\bigr],
\end{align}
\end{subequations}
where $q_{z}^{(o)}=\sqrt{n_o^2-q_x^2}$,
$q_{z}^{(e)}=n_o n_{\parallel}^{-1}\sqrt{n_{\parallel}^2-q_x^2}$
and $\delta_{\alpha}=q_{z}^{(\alpha)} k_{\vac} d$.

%%%%%%%%%%%%%%%%%%%%%%%%%%%%%%%%%%
\section{Angular structure of polarization ellipse field}
\label{sec:angular-struct}
%%%%%%%%%%%%%%%%%%%%%%%%%%%%%%%%%

In the previous section we deduced the expression for
the transmission matrix of the NLC cell and now we pass on
to discussing how the polarization properties of the transmitted
light depend on the direction of the incident wave.

This direction is specified by two angles: the incidence angle $\theta_{\inc}$
and the azimuthal angle of the plane of incidence $\phi_{\inc}$.
Clearly, we need to replace
the director azimuthal angle $\phi_{\dd}$ with
$\phi_{\dd}-\phi_{\inc}$ so as to have the Euler angles
describing orientation of the director~\eqref{eq:director} 
with respect to the incidence plane.
So, the transmission matrix
$\mvc{T}(\theta_{\inc},\theta_{\dd},\phi_{\dd})$
obtained in Sec.~\ref{sec:transm-light}
is changed to
$\mvc{T}(\theta_{\inc},\theta_{\dd},\phi_{\dd}-\phi_{\inc})$.
Dependence of the polarization parameters of transmitted waves
on the angles $\theta_{\inc}$ and $\phi_{\inc}$ will be of our primary
concern.

We begin with introducing necessary parameters 
and notations. Much of this material can be found in standard
textbooks such as~\cite{Born,Azz:1977}
(see, e.g., Ref.~\cite{Dennis:optcom:2002}
for a more recent and extended discussion). 

The polarization state of a plane wave can be conveniently
characterized by using the components of its vector amplitude
in the spherical basis as follows
\begin{subequations}
\begin{align}
&
  \label{eq:spheric-basis}
    \vc{E}(\uvc{k})=
  E_{+} \vc{e}_{+}(\uvc{k})+
  E_{-}\vc{e}_{-}(\uvc{k})
=\ee^{i\phi_0}
\{\vc{p}_{+}+i\vc{p}_{-}\},
\\
&
\label{eq:E_pm_c}
E_{\pm} =2^{-1/2}
\bigl(
E_{\parallel} \mp i E_{\perp}
\bigr)
=\bigl|E_{\pm}\bigr|
\exp\{i\phi_{\pm}\},
\\
&
\label{eq:p_pm}
\vc{p}_{\pm}= 2^{-1/2}
\bigl\{
|E_{+}|\pm |E_{-}|
\bigr\}\,
\vc{e}_{x,\,y}^{\,\prime}(\uvc{k}),
\quad
\phi_0=(\phi_{-}+\phi_{+})/2,
\end{align}
\end{subequations}
where
$\sqrt{2}\,\vc{e}_{\pm}(\uvc{k})= \vc{e}_x(\uvc{k})
\pm i \vc{e}_y(\uvc{k})$; 
the angle $\phi_0$
is variously known as 
the \textit{phase of the vibration}~\cite{Nye:prsl:1983a}
or the \textit{rectifying phase}~\cite{Dennis:optcom:2002};
$\vc{p}_{+}$ and $\vc{p}_{-}$ are
the major and minor semiaxes of the polarization ellipse
with the principal axes directed along the unit vectors
$
\vc{e}_{x}^{\,\prime}(\uvc{k})=
\cos\phi_{p}\, \vc{e}_x(\uvc{k})
+\sin\phi_{p}\, \vc{e}_y(\uvc{k})
$
and
$
\vc{e}_{y}^{\,\prime}(\uvc{k})=
-\sin\phi_{p}\, \vc{e}_x(\uvc{k})
+\cos\phi_{p}\, \vc{e}_y(\uvc{k})
$, respectively.
So, the angle
\begin{align}
  \label{eq:azim-elli}
  \phi_{p}=(\phi_{-}-\phi_{+})/2=2^{-1}\arg(\cnj{[E_{+}]} E_{-})
=2^{-1}\arg(E_{-}/E_{+})
\end{align}
specifies the orientation of the polarization ellipse and 
will be referred to as the 
\textit{azimuthal angle of polarization}
or the \textit{polarization azimuth}.

Another important characteristics of the polarization ellipse
describing its eccentricity
is the signed ellipticity parameter
\begin{align}
  \label{eq:ellipt-param}
  \epsilon_{\ellpt}=
\frac{|E_{-}|-|E_{+}|}{|E_{-}|+|E_{+}|}.
\end{align}
This parameter will be referred to as the \textit{ellipticity}
and its sign defines the handedness of the ellipse.

According to Ref.~\cite{Born}, the ellipse is considered to be
right-handed (RH) if 
its helicity is negative, $\sca{\vc{p}_{+}\times\vc{p}_{-}}{\uvc{k}}<0$,
so that $|E_{+}|<|E_{-}|$ and $\epsilon_{\ellpt}>0$.
For the left-handed (LH) ellipse,
$\sca{\vc{p}_{+}\times\vc{p}_{-}}{\uvc{k}}>0$
and $\epsilon_{\ellpt}<0$.

Experimentally, the characteristics of the polarization ellipse
can be obtained by measuring the Stokes parameters
related to 
the \textit{coherence matrix} with the elements
$\mvc{M}_{\alpha \beta}=E_{\alpha} \cnj{E_{\beta}}$,
where $\alpha,\beta\in\{\parallel,\perp\}$.
In circular basis, this matrix can be written
as a linear combination of the Pauli matrices.
The coefficients of the combination
are the Stokes parameters
\begin{align}
  \label{eq:coher-matr}
  \mvc{M}_c={\mvc{C}}\mvc{M}\hcnj{\mvc{C}}=
\begin{pmatrix}
|E_{+}|^2 & E_{+} \cnj{E}_{-}\\
E_{-} \cnj{E}_{+} & |E_{-}|^2
\end{pmatrix}
= 2^{-1}\sum_{i=0}^4
S_i\,\bs{\sigma}_i,
\end{align}
where 
$\mvc{C}=2^{-1/2}\begin{pmatrix}1 & -i \\ 1 & i\end{pmatrix}$,
$\bs{\sigma}_0\equiv\mvc{I}_2$ and 
$\bs{\sigma}_2=\begin{pmatrix}0 & -i \\ i & 0\end{pmatrix}$.
Since the determinant of the coherence matrix 
vanishes, $\det\mvc{M}=0$,
the Stokes parameters lie on the four-dimensional cone
$S_0^2=\sum_{i=1}^3 S_i^2$,
and can be parameterized as follows
\begin{subequations}
  \label{eq:Stokes}
\begin{align}
&
  \label{eq:S_0}
  S_0 = |E_{+}|^2+|E_{-}|^2=|E_{\parallel}|^2+|E_{\perp}|^2,
\\
&
\label{eq:S_1}
  S_1 = 2\Re\cnj{E}_{+} E_{-}=|E_{\parallel}|^2-|E_{\perp}|^2
=S_0\cos 2\chi_{p}\cos 2\phi_{p},
\\
&
\label{eq:S_2}
  S_2 = 2\Im\cnj{E}_{+} E_{-}=2\Re{E}_{\perp} \cnj{E}_{\parallel}
=S_0\cos 2\chi_{p}\sin 2\phi_{p},
\\
&
 \label{eq:S_3}
  S_3 =|E_{+}|^2-|E_{-}|^2 =2\Im{E}_{\perp} \cnj{E}_{\parallel}
=S_0\sin 2\chi_{p},
\end{align}
\end{subequations}
where $0<\phi_p\le\pi$ is the polarization azimuth~\eqref{eq:azim-elli}
and $-\pi/4 \le\chi_p\le \pi/4$ is the ellipticity angle.
Then, the relations expressing the ellipse characteristics
in terms of the Stokes parameters are 
\begin{align}
&
  \label{eq:ell-az-S12}
  \phi_p=2^{-1}\arg{S},\quad S\equiv S_1+ i S_2,
\\
&
\label{eq:ellip-param-S3}
\epsilon_{\ellpt}=-\tan\chi_p,\quad
\chi_p = 2^{-1}\arcsin(S_3/S_0).
\end{align}
Similar to Refs.~\cite{Dennis:optcom:2002,Jacks:bk:1999},
we have used Eq.~\eqref{eq:coher-matr} and Eq.~\eqref{eq:S_3}
to define the Stokes parameter $S_3$ which is opposite in sign
to that given in the book~\cite{Born}. 

The important special case occurs
when the wave is circularly polarized and $|E_{\nu}|=0$,
so that the phases $\phi_{\nu}$ and $\phi_p$
are indeterminate. This is an example of 
the \textit{polarization singularity} that,
according to Eq.~\eqref{eq:ell-az-S12}, 
can be regarded as the \textit{phase singularity}
of the complex Stokes field, $S=S_1+i S_2$.
The point where
the polarization is circular and $|E_{\nu}|=0$ will be referred
to as  the \textit{C$_{\nu}$ point} with $\epsilon_{\ellpt}=\nu$.
In our case such points are characterized by
the incidence angles at which the transmitted wave is circular
polarized and $E_{\nu}^{(\trans)}=0$.

The case of linearly polarized wave with $|E_{+}|=|E_{-}|$
provides another example of the polarization singularity
where the handedness is undefined.
The curves along which the polarization is linear are
called the \textit{L lines}.

\subsection{Angular patterns: C points and L lines}
\label{subsec:c-points}

Similar to Eq.~\eqref{eq:transm-rel}, 
the transmission matrix
in circular basis
\begin{align}
  \label{eq:T-circul}
  \mvc{T}_c=
\begin{pmatrix}
t_{++} & t_{+-}\\ t_{-+} & t _{--}
\end{pmatrix}
={\mvc{C}}\mvc{T}\hcnj{\mvc{C}}
\end{align}
relates the circular components of the incident and transmitted waves,
$\{E_{+}^{(\inc)}, E_{-}^{(\inc)}\}$ and $\{E_{+}^{(\trans)}, E_{-}^{(\trans)}\}$.
Since the transmission matrix~\eqref{eq:Tmatr-expl}
is symmetric, the diagonal elements of the matrix~\eqref{eq:T-circul}
are equal, $t_{++}=t_{--}=(t_{xx}+t_{yy})/2$, whereas the non-diagonal
elements $t_{\pm\mp}=(t_{xx}-t_{yy})/2\mp t_{xy}$ differ
provided $\mvc{T}$ is not diagonal. 

The transmission matrix
\begin{align}
&
  \label{eq:tilde-T}
  \tilde{\mvc{T}}(\rho,\phi)
=\exp(-i\phi\,\bs{\sigma}_3)\mvc{T}_c(\rho,\phi)\exp(i\phi\,\bs{\sigma}_3),
\\
&
\label{eq:rho}
\rho= r \tan\theta_{\inc},\quad
\phi= \phi_{\inc}
\end{align}
describes the conoscopic patterns
on the transverse plane of projection, 
where $\rho$ and $\phi$ are the polar coordinates
(the Cartesian coordinates are:
$x=\rho\cos\phi$ and $y=\rho\sin\phi$)
and 
$r$ is the aperture dependent scale factor.
We concentrate on the case,
in which the incident plane wave is linearly polarized
along the unit vector:
$
\cos\psi_{p}\, \vc{e}_x(\uvc{k}_{\inc})
+\sin\psi_{p}\, \vc{e}_y(\uvc{k}_{\inc})
$.
For the incident wave with the circular components 
$E_{\nu}^{(\inc)}=\exp(-\nu\psi_{p}) |E_{\inc}|$,
the reduced components of the transmitted wave are given by
\begin{align}
  \label{eq:E_mu_tr}
  E_{\nu}^{(\trans)}/|E_{\inc}|\equiv\Psi_{\nu}/2=
\bigl[
t_{\nu,\,\nu} + t_{\nu,\,-\nu}\exp(-2i\nu\psi)
\bigr]
\exp(-i\nu\psi_p),
\end{align}
where $\psi=\phi-\psi_p$.

The polar coordinates of the C$_{\nu}$ points 
on the projection plane,
$\rho_{k}^{(\nu)}$ and $\phi_{k}^{(\nu)}$,
where $k$ is the numbering label,
can be found by solving the equation
\begin{align}
  \label{eq:C-mu-gen}
  |\Psi_{\nu}(\rho,\phi)|=0
\end{align}
that generally has multiple solutions.

The C$_{\nu}$ points can be viewed as the phase
singularities of the complex scalar field
\begin{align}
  \label{eq:tld-S}
  \tilde{S}=\cnj{\Psi}_{+}\Psi_{-}=\tilde{S}_1+i\tilde{S}_2
\end{align}
proportional to the Stokes field defined in Eq.~\eqref{eq:ell-az-S12}.
Such singularities are characterized by 
the \textit{winding number} which is the signed number
of rotations  of the two-component field
$(\tilde{S}_1,\tilde{S}_2)$ around the circuit surrounding the
singularity~\cite{Merm:rpm:1979}.
The winding number also known as the \textit{signed strength of the
dislocation} is generically $\pm 1$.

Since the polarization azimuth~\eqref{eq:ell-az-S12} is defined modulo
$\pi$ and $2\phi_p=\arg\tilde{S}$, the dislocation strength is twice
the index of the corresponding C$_{\nu}$ point, $I_{C}$.
For generic C points, $I_{C}=\pm 1/2$
and the index can be computed from the formula
\begin{align}
  \label{eq:I_c-gen}
  I_{C}=\frac{1}{2}\sign
\Bigr[
\Im(\pdrs{x}\cnj{\tilde{S}}\pdrs{y}\tilde{S})
\Bigr]_{\substack{x=x_{\nu}\\ y=y_{\nu}}},
\end{align}
where $\pdrs{x} f$ is the partial derivative of $f$ with respect to $x$.

The relation~\eqref{eq:I_c-gen} gives the index of the C$_{\nu}$ point with
the coordinates $(x_{\nu},y_{\nu})$
expressed in terms of the vorticity~\cite{Dennis:prsl:2000,Dennis:optcom:2002}:
$\Im(\pdrs{x}\cnj{\tilde{S}}\pdrs{y}\tilde{S})
=\pdrs{x}\tilde{S}_1\pdrs{y}\tilde{S}_2-
\pdrs{y}\tilde{S}_1\pdrs{x}\tilde{S}_2
$.
The formula linking gradients of
the complex field 
$\sca{\vc{E}}{\vc{E}}\propto\Psi_{+}\Psi_{-}$
and the index for C lines in the three-dimensional space
was derived in Ref.~\cite{Berry:jopa2:2004}.

If $\Psi_{\nu}=0$ (C$_{\nu}$ point), 
only derivatives of $\Psi_{\nu}$ enter the
expression~\eqref{eq:I_c-gen}
which can be suitably rearranged to yield 
the index of the C$_{\nu}$ point in the following form:
\begin{align}
  \label{eq:I_c-Psi_mu}
  I_{C}=\frac{\nu}{2}\sign
\Bigr[
\Im(\pdrs{x}\Psi_{\nu}\,\pdrs{y}\cnj{\Psi}_{\nu})
\Bigr]_{\substack{x=x_{\nu}\\ y=y_{\nu}}}
=
\frac{\nu}{2}\sign
\Bigr[
\Im(\pdrs{\rho}\Psi_{\nu}\,\pdrs{\phi}\cnj{\Psi}_{\nu})
\Bigr]_{\substack{\rho=\rho_{\nu}\\ \phi=\phi_{\nu}}}.
\end{align}
Subsequently, we shall apply
the formula~\eqref{eq:I_c-Psi_mu} expressing the index in terms of
the derivatives with respect to polar coordinates 
to the case of homeotropically aligned cell.

In addition to the handedness and the index,
the C points are classified according to the number of 
straight lines terminating on the singularity.
This is the so-called \textit{line classification}
that was initially studied in the context of 
umbilic points~\cite{Berry:jpa:1977}.

For generic C points, the number of the straight lines, $N_{C}$,
may either be 1 or 3.
This number is $3$ provided the index equals $-1/2$, $I_{C}=-1/2$,
and such C points are called \textit{stars}.
At $I_{C}=1/2$, there are two characteristic patterns of polarization ellipses
around a C point: (a)~\textit{lemon} with $N_{C}=1$ and 
(b)~\textit{monstar} with $N_{C}=3$~\cite{Nye:prsl:1983a}.
The quantitative criterion to distinguish between the C points of
the lemon and the monstar types was deduced
in~\cite{Dennis:optcom:2002}.

We conclude this subsection with the remark that 
the transmitted wave is linearly polarized
when the condition
\begin{align}
  \label{eq:L-lines}
  |\Psi_{+}(\rho,\phi)|=|\Psi_{-}(\rho,\phi)|
\end{align}
is satisfied.
So, Eq.~\eqref{eq:L-lines} describes loci of points forming
the L lines lying in the projection plane.

\begin{figure*}[tbh]
%\vskip5mm
\centering
\subfigure[Homeotropic structure]{%
\resizebox{65mm}{!}{\includegraphics*{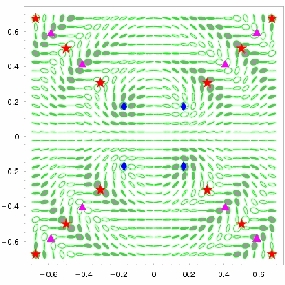}}
\label{fig:hom-th}
}
\subfigure[Tilted structure at $\theta_{\dd}=30$~deg]{%
\resizebox{65mm}{!}{\includegraphics*{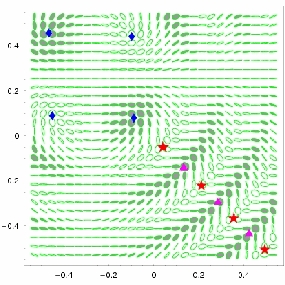}}
\label{fig:tlt70-th}
}
\subfigure[Tilted structure at $\theta_{\dd}=60$~deg]{%
\resizebox{65mm}{!}{\includegraphics*{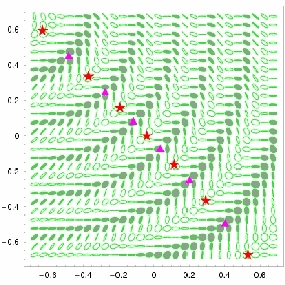}}
\label{fig:tlt30-th}
}
\subfigure[Planar structure]{%
\resizebox{65mm}{!}{\includegraphics*{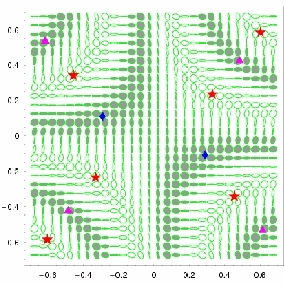}}
\label{fig:planar-th}
}
\caption{%
The angular structure of polarization 
computed as the polarization ellipse field in the projection plane
where the polar coordinates are $\rho=\tan\theta_{\inc}$
and $\phi=\phi_{\inc}$.
The incident light of the wavelength
633~nm is linearly polarized with $\psi_p=0$.
Other parameters used in calculations are: 
the cell thickness is $d=18.7$ ~$\mu$m;
$\mu =\mu_{\med}=1$;
$n_{\med}=1.5$;
$n_e=n_1=1.71$ and $n_o=n_2=n_3=1.527$ (5CB).
The C points of the lemon and
the monstar types are marked by diamonds and triangles, respectively.
Stars are used for the C points of the star type.
Left-handed and right-handed
polarization is respectively indicated by 
open and filled ellipses. 
}
\label{fig:theory-uniax}
\end{figure*}

\begin{figure*}[tbh]
%\vskip5mm
\centering
\subfigure[Biaxial structure at $\theta_{\dd}=0$~deg.]{%
\resizebox{65mm}{!}{\includegraphics*{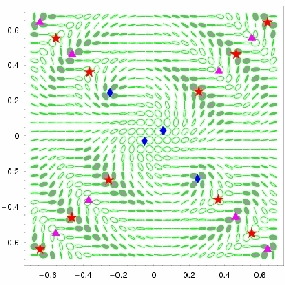}}
\label{fig:biax-hom}
}
\subfigure[Biaxial structure at $\theta_{\dd}=10$~deg.]{%
\resizebox{65mm}{!}{\includegraphics*{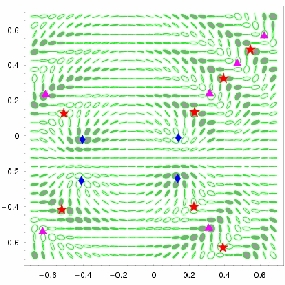}}
\label{fig:biax-tlt}
}
\caption{%
The angular structure of polarization 
computed as the polarization ellipse field in the projection plane
where the polar coordinates are $\rho=\tan\theta_{\inc}$
and $\phi=\phi_{\inc}$.
The incident light of the wavelength
633~nm is linearly polarized with $\psi_p=0$.
Other parameters used in calculations are: 
the cell thickness is $d=18.7$ ~$\mu$m;
$\mu$ =1; $\gamma_{\dd}=45$~degrees;
$n_1=1.71$, $n_2=1.52$ and $n_3=1.53$.
The C points of the lemon and
the monstar types are marked by diamonds and triangles, respectively.
Stars are used for the C points of the star type.
Left-handed and right-handed
polarization is respectively indicated by 
open and filled ellipses. 
}
\label{fig:theory-biax}
\end{figure*}

% \begin{figure*}[tbh]
% %\vskip5mm
% \centering
% \subfigure[Theory]{%
% \resizebox{75mm}{!}{\includegraphics*{th_hom_90r.eps}}
% \label{fig:hom-th-90}
% }
% \subfigure[Experiment]{%
% \resizebox{75mm}{!}{\includegraphics*{exp_hom_90r.eps}}
% \label{fig:hom-exp-90}
% }
% \caption{%
% The angular structure of polarization at $\phi_p=\pi/2$
% shown as the polarization ellipse field in the projection plane
% where the polar coordinates are $\rho=r \tan\theta_{\inc}$
% and $\phi=\phi_{\inc}$.
% }
% \label{fig:hom-exp-th-90}
% \end{figure*}

\subsection{Homeotropic cell}
\label{subsec:homeot-cell}

When the director is normal to the substrates,
the NLC cell is homeotropically aligned and $d_z=1$.
In this case the transmission matrix is diagonal
and the circular components of the transmitted light
are given by
\begin{align}
  \label{eq:Psi_mu}
  \Psi_{\nu}=
\bigl[
(t_{x}+t_{y}) + (t_x-t_y)\exp(-2i\nu\psi)
\bigr]
\exp(-i\nu\psi_p),
\end{align}
where the transmission coefficients $t_{x,\,y}=t_{x,\,y}(\theta_{\inc})\equiv t_{x,\,y}(\rho)$
do not depend on the azimuthal angle of the incidence plane
$\phi\equiv\phi_{\inc}$
and are defined by Eq.~\eqref{eq:giag-transm} and Eq.~\eqref{eq:tau-homeotr}.

From Eq.~\eqref{eq:Psi_mu}, the C points may appear at the incidence angles $\theta_{\inc}$
that satisfy the condition: $|t_x+t_y|=|t_x-t_y|$.
Another form of this condition
\begin{align}
  \label{eq:C-mu-cond}
  R(\rho)\equiv\Re(t_x\cnj{t}_y)=|t_x|\cdot |t_y|\cos\delta=0,
\end{align}
where $\delta$ is the phase difference, defines the radii of circles
containing the C points.
Note that, in weakly anisotropic NLCs, 
the phase difference $\delta$ can be approximated by
the phase shift due to the difference in optical path
of the ordinary and extraordinary waves,
$\delta\approx\delta_e-\delta_o$,
for the incidence angles up to $70$ degrees.
Clearly, Eq.~\eqref{eq:C-mu-cond} implies that $\delta=\pi/2+\pi k$,
where $k=0, 1\ldots N-1$ is the non-negative integer 
and $N$ is the number of solutions.
So, in the transverse projection plane,
there are $N$ circles with C points. 
The radii of the circles $\rho_k$, $k=0, 1\ldots N-1$,
can be found by solving Eq.~\eqref{eq:C-mu-cond}. 

At $\rho=\rho_k$, 
it is not difficult to obtain the expressions for
the amplitude of the components~\eqref{eq:Psi_mu}
\begin{align}
&
  \label{eq:Psi2_mu-rhok}
  |\Psi_{\nu}(\rho_k,\phi)|^2= 2\, |t|^2
\bigl[
1+\cos 2\{\psi-\nu (-1)^k\alpha\}
\bigr],
\\
&
\label{eq:alpha}
|t|^2 = |t_x|^2+|t_y|^2,\quad
\tan\alpha=\frac{|t_y|}{|t_x|}.
\end{align}
From Eq.~\eqref{eq:Psi2_mu-rhok}, it immediately follows
that there are two pairs of C points with 
the azimuthal angles given by
\begin{align}
  \label{eq:phi-mu}
  \phi_{\pm k}^{(\nu)}=\psi_p\pm\pi/2+\nu (-1)^{k}\alpha
\end{align}
in each circle of the radius $\rho_k$.
The symmetric arrangement of the C points 
[see also Fig.~\ref{fig:hom-th})]
is a consequence of the symmetry relations
\begin{align}
  \label{eq:symmetry-homeo}
  |\Psi_{\nu}(\rho,\psi)|=|\Psi_{\nu}(\rho,\pi+\psi)|,
\quad
  |\Psi_{\nu}(\rho,\psi)|=|\Psi_{-\nu}(\rho,\pi-\psi)|
\end{align}
for the amplitudes~\eqref{eq:Psi_mu}.

We can now substitute Eq.~\eqref{eq:Psi_mu} into the
expression for the index~\eqref{eq:I_c-Psi_mu}
to derive the result
\begin{align}
  \label{eq:I_c-homeo}
  I_{C}=-\frac{1}{2}\,\pdrs{\rho} R(\rho)\Bigr|_{\rho=\rho_k}=\frac{(-1)^k}{2},
\end{align}
where the second equality follows because
$R$ is an oscillating function of $\rho$ and $R(0)>0$.

Interestingly, Eq.~\eqref{eq:I_c-homeo} relates the index of C points
and derivatives of the transmission coefficients with
respect to the incidence angle 
($\pdrs{\rho} f =\cos^2\theta_{\inc}\,\partial
f/\partial\theta_{\inc}$).
The index is thus determined by the circle number $k$
and alternates in sign starting from $I_{C}=+1/2$.
So, the C points in the vicinity of the origin at $\rho=\rho_0$ 
can either be lemons or monstars.

Now it is our task to formulate the criterion
to distinguish between these two types.
By using Eq.~\eqref{eq:Psi_mu} we find
the complex Stokes field~\eqref{eq:tld-S} and the polarization
azimuth~\eqref{eq:ell-az-S12}
\begin{align}
&
  \label{eq:tld-S-homeo}
  \tilde{S}= 4\ee^{2i\phi} P,
\quad 
P=|t_x|^2\cos^2\psi-|t_y|^2\sin^2\psi-i R\,\sin 2\psi,
\\
&
\label{eq:phi_p-homeo}
2\phi_p = 2\phi+\arg P
\end{align}
for the homeotropic cell.

Our next step is to consider 
the first order term of the power series expansion
of $P$ in $\Delta x=\Delta r\cos\phi_r$
and $\Delta y=\Delta r \sin\phi_r$.
This is the linear part of the expansion 
that can be conveniently expressed in polar coordinates as follows
\begin{align}
&
  \label{eq:Delta-P}
  \rho\Delta P\approx
\rho
\bigl[
\Delta x\,\pdrs{x} P+\Delta y\,\pdrs{y} P
\bigr]
=
\notag
\\
&
\Delta r
\bigl[
\cos(\phi_r-\phi)
\rho\,
\pdrs{\rho} P+
\sin(\phi_r-\phi)\,\pdrs{\phi} P
\bigr].
\end{align}
At a C point, the expression~\eqref{eq:Delta-P}
represents the complex field $P$ expanded to first order. 
The identity 
\begin{align} 
 \label{eq:Re-P}
  \Re P =|t|^2\cos(\psi-\alpha)\cos(\psi+\alpha).
\end{align}
eases computing the derivatives at the C point.
The result is
\begin{align}
&
  \label{eq:DP-phi}
\pdrs{\phi} P\Bigr|_{\substack{\rho=\rho_k\\ \psi=\pi/2\pm\alpha}}
=\pm |t|^2\sin(2\alpha),
\\
&
  \label{eq:DP-rho}
\rho\pdrs{\rho} P\Bigr|_{\substack{\rho=\rho_k\\ \psi=\pi/2\pm\alpha}}
=\sin(2\alpha)
\bigl[
-|t|^2\alpha_1 \pm i R_1
\bigr],
\\
&
\label{eq:R1-alpha1}
R_1\equiv (\rho\pdrs{\rho} R)\Bigr|_{\rho=\rho_k},
\quad
\alpha_1\equiv (\rho\pdrs{\rho} \alpha)\Bigr|_{\rho=\rho_k}.
\end{align}
We can now follow the line of reasoning presented in
Ref.~\cite{Dennis:optcom:2002} and derive the polynomial
equation
\begin{align}
  \label{eq:L-eq}
  \tan 2(\phi_p-\phi)=\frac{2 q}{1-q^2}=
\frac{R_1}{|t|^2 (q\mp\alpha_1)}=
\frac{\Im\Delta P}{\Re\Delta P}
\end{align}
in $q\equiv\tan(\phi_p-\phi)=\tan(\phi_r-\phi)$.
The roots of Eq.~\eqref{eq:L-eq} 
give the angles where $\phi_p=\phi_r$
and their number depends on the sign of the discriminant
\begin{align}
  \label{eq:D-L}
  D_{L}=(R_1/|t|^2+1)^2+\alpha_1^2-1.
\end{align}

For $D_L<0$, $N_{C}=1$ and the point is of the lemon type.
In the opposite case with $D_{L}>0$, the C points can either be
stars or monstars depending on the index, 
$I_{C}=-1/2$ for stars and $I_{C}=1/2$ for monstars.

As it can be seen from Eq.~\eqref{eq:I_c-homeo},
the derivative $R_1$ is positive if $I_{C}=-1/2$
and, as a result, $D_L>0$. So, in agreement with
the known result, our criterion leads to the conclusion 
 that the C points with $I_{C}=-1/2$ are always stars.

At $I_{C}=1/2$, $R_1$ is negative
and the point will be monstar provided the gradients
$|R_1|$ and $|\alpha_1|$ are sufficiently large.
For example, the points where $R_1<-2|t|^2$ are monstars.
As is illustrated in Fig.~\ref{fig:hom-th}, 
in sufficiently thick cells, the C points with $k\ge 2$ and $I_{C}=1/2$
are monstars, whereas the C points in the circle of the smallest
radius ($k=0$) are lemons.

More generally, monstars will be the dominating type in regions of
pronounced dependence of the transmission coefficients on the
incidence angle.
By contrast, a mechanism suppressing strong inhomogeneity
may drastically increase the fraction of lemons.
For example, in liquid crystals,
strong director deformations (and large gradients)
are  inhibited due to the elastic energy costs 
and, thus, disclinations of the monstar type are generally unstable~\cite{Gennes:bk:1993}.  

For the homeotropic configuration, the condition~\eqref{eq:L-lines}
can be simplified giving the equation 
\begin{align}
  \label{eq:L-lines-homeo}
  \Im(t_x \cnj{t}_y)\sin 2\psi=0
\end{align}
that describes the L lines. 
From Eq.~\eqref{eq:L-lines-homeo} there are two
straight lines of linear polarization: $\phi=\psi_p$ and
$\phi=\psi_p+\pi/2$, where 
the polarization vectors of incident and transmitted waves are parallel,
$\phi_p=\psi_p$. 

Other L lines are circles separating the circles of C
points. 
The radii of the circles, $\rho_{k}^{(L)}$, can be found as the solutions
to the equation $\delta=\pi k$ with $k=1,\ldots N-1$.
When $k$ is even and $|t_x|\approx |t_y|$,
from Eq.~\eqref{eq:phi_p-homeo} it can be concluded that,
similar to the straight L lines, $\phi_p\approx \psi_p$.
If $k$ is odd, the polarization vector rotates with the azimuthal
angle of the incidence plane and $\phi_p\approx \psi_p+2\phi$.
These can also be seen in Fig.~\ref{fig:hom-th}. 

\subsection{Director tilt and biaxiality induced effects}
\label{subsec:director-unfold}

The analytical results presented in
the previous section completely characterize the transmission angular pattern
of the light polarization for the homeotropic orientational structure
in terms of the polarization singularities.
In Fig.~\ref{fig:hom-th}, this polarization resolved conoscopic
pattern is depicted as the field of polarization ellipses
in the projection plane where $x=\tan\theta_{\inc}\cos\phi_{\inc}$
and $y=\tan\theta_{\inc}\sin\phi_{\inc}$.

Referring to Fig.~\ref{fig:hom-th} [see also Eq.~\eqref{eq:phi-mu}], 
when $|t_x|\approx |t_y|$ and
$\alpha\approx\pi/4$,
the C points are arranged in chains formed by four rays
along which they alternate in sign of the handedness
and of the index.
The symmetry of the chain structure can be deduced from the symmetry 
relations~\eqref{eq:symmetry-homeo}.

In particular, the structure is invariant with respect to 
the operation of reflection: $\psi\to -\psi$ and $\nu\to -\nu$.
It additionally posses the symmetry center located at the
origin and associated with the transformation of inversion: $\phi\to
\phi+\pi$. Clearly, each of the two straight L lines, $\phi=\psi_p$ and
$\phi=\psi_p+\pi/2$, is the symmetry axis and the symmetry
center is also the center of the L circles.

Algebraically, the symmetry properties of the polarization pattern
follow because the transmission matrix
for the homeotropic cell~\eqref{eq:giag-transm} 
is diagonal and the transmission coefficients,
$t_x$ and $t_y$, are independent of the azimuthal angle
of the incidence plane, $\phi=\phi_{\inc}$.
For tilted director configurations with
$0<\theta_{\dd}<\pi/2$, this is no longer the case.

In this case, the transmission matrix~\eqref{eq:Tmatr-expl} is 
non-diagonal and its elements depend on $\phi_{\dd}-\phi$.
Generally, there are two types of the azimuthal angle dependent terms:
(a)~the terms proportional to $\cos 2(\phi_{\dd}-\phi)$
and (b)~the inversion symmetry breaking terms
proportional to either $\sin(\phi_{\dd}-\phi)$ or $\cos(\phi_{\dd}-\phi)$.

In Fig.~\ref{fig:theory-uniax}, we show what happen when
the tilt angle $\theta_{\dd}$ varies from $\theta_{\dd}=0$ to
$\theta_{\dd}=\pi/2$, so that the orientational structure changes
from the homeotropic configuration to the planar one.
From Fig.~\ref{fig:tlt70-th} it is seen that, at small incidence angles, 
the symmetry breaking terms result in the shift of the symmetry
center. When the incidence angle $\theta_{\dd}$ increases further,
the center of symmetry eventually leaves the region specified by the aperture
leading to the single-chain structure of the C points shown in
Fig.~\ref{fig:tlt30-th}.

In the case of the planar structure with $d_z=0$
($\theta_{\dd}=\pi/2$), it is not difficult to see that
the terms dependent on the azimuthal angle $\phi$
are proportional either to $d_y^2$ or to $d_x^2$.
So, similar to the homeotropic cell,
the origin is the symmetry center for
the polarization pattern of light passed through the planar cell.
Fig.~\ref{fig:planar-th} demonstrates that, by contrast to the
homeotropic structure, where the polarization field is of the
elliptic type (see Fig.~\ref{fig:hom-th}), 
the angular pattern of polarization is of the hyperbolic type
for the planar structure.     

The results computed for weakly biaxial orientational configurations
are presented in Fig.~\ref{fig:theory-biax}.
In these calculations, the Euler angles, $\phi_{\dd}$ and
$\gamma_{\dd}$,
that specify orientation of the principal axes~\eqref{eq:director} are
taken to be zero and $\pi/4$, respectively.
In Fig.~\ref{fig:biax-hom}, similar to the homeotropic cell, 
the tilt angle $\theta_{\dd}$ is zero,
whereas the polarization
pattern for the tilted biaxial structure [see Fig.~\ref{fig:biax-tlt}]
was calculated at $\theta_{\dd}=10$ degrees.
From the patterns plotted in Fig.~\ref{fig:biax-hom} and
Fig.~\ref{fig:biax-tlt} it is clear that biaxiality induces
additional deformations of the polarization ellipse field
as compared to the case of uniaxially anisotropic cells.

%%%%%%%%%%%%%%%%%%%%%%%%
\section{Discussion and conclusions}
\label{sec:disc-concl}
%%%%%%%%%%%%%%%%%%%%%%%

In this paper we have studied the polarization resolved
angular patterns of light transmitted through uniformly anisotropic
NLC cells. Such patterns can be considered 
as the polarization structure underlying 
the conoscopic images measured
in experiments with two crossed polarizers.

Experimentally, this structure can be investigated
using a suitably modified method of the Stokes polarimetry 
designed for accurate measurements of  the Stokes 
parameters (typical experimental setups are described, 
e.g., in Refs.~\cite{Soskin:optl:2003,Dennis:prl:2005}).
The predictions of our theory for the homeotropic cell
was successfully compared with the polarization patterns
experimentally obtained by the group of experimentalists
at Institute of Physics of NASU (Kyiv, Ukraine).
These results are the subject of a separate joint publication.

The exact solution to the transmission (reflection) problem 
for the uniformly anisotropic cell was derived   
in Sec.~\ref{sec:transm-light} by using the $4\times 4$ matrix
formalism in combination with the orthogonality relations.
We deduced the relation linking the operator of evolution
and the transmission (reflection) matrix
[see Eq.~\eqref{eq:trans-mat1} and Eq.~\eqref{eq:refl-mat1}].
Then the basis of eigenwaves was used to construct 
the operator of evolution~\eqref{eq:oprt-evol-nlc}.
After simplifying algebraic rearrangements
made with the help of the orthogonality relations~\eqref{eq:orth-matr},
the final result
[see Eq.~\eqref{eq:Tmatr-expl} and Eq.~\eqref{eq:Rmatr-expl}]
was expressed in terms of $2\times 2$ matrices.

For the case of one interfacial (reflecting) surface, the explicit formulas
in the $2\times 2$ matrix form
were previously derived in Ref.~\cite{Lekner:jpcm:1991,Lekner:1:jpcm:1992}.
Algorithms and useful approximations for inhomogeneous $z$-stratified
anisotropic media were suggested in more recent 
papers~\cite{Yakovl:opsp:1999,Yakovl:opsp:2003,Yakovl2:opsp:2003,Palto:jetp:2001}.
For our purposes, however, it is suffice to have the exact
expressions in the compact convenient form that emphasize the symmetry
properties of the transmission matrix and enormously simplifies numerical
calculations. An extended discussion of generalizations and
approximation schemes is beyond the scope of this paper.

The analytical results  
were applied to characterize the polarization resolved angular patterns
in terms of the polarization singularities such as C points and L
lines. The case of the homeotropic director configuration
was analyzed in detail. We found that the singularities are
symmetrically arranged and deduced the simple formula for the index of
C points. We have also formulated the quantitative criterion to
determine the type of the C points.

The C points are shown to form chains along which
they alternate in sign of the handedness and the index, $I_{C}$.
At $I_{C}=-1/2$, all the C points are stars, whereas the C points
with $I_{C}=+1/2$ can either be lemons or monstars depending
on how strong the dependence of the transmission coefficients
on the incidence angle is.
In sufficiently thick NLC layers ($d>10$~\mum\, for 5CB),
the C points of the lemon type are found to be located in the immediate vicinity
of  the origin.  

The patterns for the tilted, planar and weakly biaxial configurations
were computed numerically and presented as the polarization ellipse fields. 
The primary goal of these calculations is 
to demonstrate the effects of  director orientation and biaxiality 
on the polarization resolved angular patterns.
At this stage we can only hope that these effects can be used to develop
an improved experimental technique for detecting the peculiarities of
liquid crystal orientational structures.  
But our results can be regarded as a first step in this direction. 

Applicability of our results is not restricted to the case
of nematic cells that were discussed as technoligically important
materials. In our considerations, a nematic liquid crystal can be easily
changed for another  uniformly anisotropic medium.

Our concluding remark concerns the effects related to 
the polarization state of the incident light.
For definiteness,
in Sec.~\ref{subsec:homeot-cell} and Sec.~\ref{subsec:director-unfold},
this wave was assumed to be linearly polarized.
More generally, 
ellipticity and polarization azimuth of the incident wave can be
regarded as additional governing parameters that may have a profound
effect on the polarization angular patterns.
For example, if the incident light is circular polarized, 
the angular pattern for the hometropic cell is characterized by
the only C point located at the origin.

The effects of the incident wave polarization can be studied in terms 
of generalized polarization patterns in differently parameterized
planes. We shall extend on this subject in our subsequent publications.

%\bibliographystyle{apsrev}
%\bibliographystyle{lc}
%\bibliography{optics,polymer,scatter,lc,quant,hk,flc,qft}

\end{document}